\documentclass[12pt]{article}
\usepackage{epsf}
\usepackage{amsfonts}
\usepackage{amssymb}
\usepackage{xcolor}
\usepackage{color}
\usepackage{graphics}
\usepackage{graphicx}
\begin{document}
\title{{\bf False vacuum as an unstable state:\\ possible cosmological implications}\footnote{A talk given at \textit{25$^{th}$ Recontres de Blois:
Particle Physics and Cosmology}, Blois,  May 26 --- 31, 2013}}
\author{K. Urbanowski\footnote{e--mail: K.Urbanowski@proton.if.uz.zgora.pl}, \\
\hfill\\
University of Zielona G\'{o}ra, Institute of Physics, \\
ul. Prof. Z. Szafrana 4a, 65--516 Zielona G\'{o}ra, Poland.}
\maketitle

\begin{abstract}
 Recent LHC results concerning the mass of the Higgs boson
 indicate that the vacuum in our Universe may be
 unstable.
 We analyze properties of  unstable
 vacuum states from the point of view of the quantum theory
 of unstable states. From the literature it is known  that some of false vacuum
 states may survive up to times when their survival probability
 has a non-exponential form. At times much latter than the
 transition time, when contributions to the survival probability
 of its exponential and non-exponential parts are comparable,
 the survival probability as a function of time $t$ has an
 inverse power--like form.  We show that at this time region
 the instantaneous energy of the false vacuum states tends
 to the energy of the true vacuum state as $1/t^{2}$ for
 $t \to \infty$.  Properties of the instantaneous energy
 at transition times are also analyzed for a given model.
 It is shown that at this time region large and rapid
 fluctuations of  the instantaneous energy take place.
 This suggests analogous behavior of the cosmological
 constant at these time regions.
\end{abstract}

\section{Introduction}

The problem of false vacuum decay became famous after the publication of
pioneer papers by Coleman and his colleagues
\cite{Coleman,Coleman2}.
The instability of a physical system in
a state which is not an absolute minimum of its energy
density, and which is separated from the minimum by
an effective potential barrier was discussed there. It was shown, in those papers,
that even if the state of the early Universe is too
cold to activate a "{\it thermal}" transition (via thermal
fluctuations) to the lowest energy (i.e. {\it "true vacuum"}) state,
a quantum decay from the false vacuum to the true vacuum
may still be possible through a barrier penetration
by macroscopic quantum tunneling.
Not long ago,
the decay of the false vacuum state in a cosmological
context has attracted
interest, in respect of possible
tunneling processes among the many vacuum states of the string
landscape (a set of vacua in the low energy
approximation of string theory).
In many models the scalar field potential driving inflation has a multiple,
low--energy minima or \textit{"false vacuua"}. Then the absolute minimum of the energy density
is the \textit{"true vacuum"}.

Krauss nad Dent analyzing a false vacuum decay \cite{Krauss}
pointed out that in eternal inflation, even though regions of false vacua by assumption
should  decay exponentially, gravitational effects force space in a region that has not decayed yet
to grow exponentially fast.
This effect causes that many false vacuum regions
can survive up to the times much later than times when the exponential decay law
holds. In the mentioned paper by Krauss and Dent the attention
was focused on the possible
behavior of the unstable false vacuum at very late times, where deviations from the exponential
decay law become to be dominat.

Recently the problem of the instability
the false vacuum state
triggered much discussion in the context of the discovery of the Higgs--like resonance at 125 --- 126 GeV (see, eg., \cite{Spencer} --- \cite{Wei}).
In the recent analysis \cite{Degrassi} assuming  the validity of the Standard Model up to Planckian energies it was shown that a Higgs mass $m_{h} < 126$ GeV implies that the electroweak vacuum  is a metastable state. This means that a discussion of Higgs vacuum stability must be considered in a cosmological framework, especially when analyzing inflationary processes or the process  of tunneling among the many vacuum states of the string
landscape.

The aim of these considerations is to analyze properties of the false vacuum state as an unstable state,  the form of the decay law  and to discuss the late time behavior of the energy of the false vacuum states.

\section{Unstable states in short}

If $|M\rangle$ is an initial unstable
state then the survival probability, ${\cal P}(t)$, equals
\[
{\cal P}(t) = |a(t)|^{2},
\]
where $a(t)$ is the survival amplitude,
\[
a(t) = \langle M|M;t\rangle,\;\;\;{\rm and}\;\; \; a(0) = 1,
\]
and
\[|M;t\rangle =
e^{\textstyle{-it\mathfrak{H}}}\,|M\rangle,
\]
$\mathfrak{H}$ is the total Hamiltonian of the system under considerations.
The spectrum, $\sigma(\mathfrak{H})$, of $\mathfrak{H}$ is assumed to be bounded from below, $\sigma(\mathfrak{H}) =[E_{min},\infty)$
and $E_{min} > -\infty$.

From basic principles of quantum theory it is known that the
amplitude $a(t)$, and thus the decay law ${\cal P}(t)$ of the
unstable state $|M\rangle$, are completely determined by the
density of the energy distribution function $\omega({\cal E})$ for the system
in this state
\begin{equation}
a(t) = \int_{Spec.(\mathfrak{H})} \omega({ E})\;
e^{\textstyle{-\,i\,{ E}\,t}}\,d{ E}.
\label{a-spec}
\end{equation}
where
\[
\omega({E}) \geq 0\;\;{\rm for} \;\;E \geq E_{min}\;\;  {\rm and} \;\;\;\;\omega ({ E}) = 0 \;\;\; {\rm for} \;\;\;E < E_{min}.
\]
From this last condition and from the Paley--Wiener
Theorem it follows that there must be \cite{Khalfin}
\[
|a(t)| \; \geq \; A\,e^{\textstyle - b \,t^{q}}, \label{|a(t)|-as}
\]
for $|t| \rightarrow \infty$. Here $A > 0,\,b> 0$ and $ 0 < q < 1$.
This means that the decay law ${\cal P}(t)$ of unstable
states decaying in the vacuum can not be described by
an exponential function of time $t$ if time $t$ is suitably long, $t
\rightarrow \infty$, and that for these lengths of time ${\cal P}(t)$ tends to zero as $t \rightarrow \infty$  more slowly
than any exponential function of $t$.
The analysis of the models of
the decay processes shows that
\[
{\cal P}(t) \simeq
e^{\textstyle{- {\it\Gamma}_{M}t}},
\]
 (where
$\Gamma_{M}$ is the decay rate of the state $|M \rangle$),
to an very high accuracy  at the canonical decay times $t$:
From $t$ suitably later than the initial instant $t_{0}$
up to
 \[
 t \gg \tau_{M} = \frac{1}{{\it\Gamma}_{M}}
 \]
($\tau_{M}$ is a lifetime) and smaller than $t = T$, where $T$ is the crossover time and denotes the
time $t$ for which the non--exponential deviations of $a(t)$
begin to dominate.

In general,
in the case of quasi--stationary (metastable) states it is convenient to express $a(t)$ in the
following form
\begin{equation}
a(t) = a_{exp}(t) + a_{non}(t), \label{a-exp+}
\end{equation}
where $a_{exp}(t)$ is the exponential part of $a(t)$, that is
\begin{equation}
a_{exp}(t) =
N\,e^{\textstyle{-it(E_{M} - \frac{i}{2}\,{\it\Gamma}_{M})}}, \label{a-exp}
\end{equation}
($E_{M}$ is the energy of the system in the state $|M\rangle$ measured at the canonical decay times,
$N$ is the normalization constant), and $a_{non}(t)$ is the
non--exponential part of $a(t)$. For times $t \sim \tau_{M}$:
\[
|a_{exp}(t)| \gg |a_{non}(t)|,
\]
The crossover time $T$
can be found by solving the following equation,
\begin{equation}
|a_{exp}(t)|^{\,2} = |a_{non}(t)|^{\,2}.
\end{equation}
The amplitude $a_{non}(t)$ exhibits inverse
power--law behavior at the late time region: $t \gg T$.
Indeed,  the integral representation (\ref{a-spec}) of $a(t)$ means that $a(t)$ is
the Fourier transform of the energy distribution function $\omega(E)$. Using this fact we can find
asymptotic form of $a(t)$ for $t \rightarrow \infty$. Results are rigorous (see  \cite{epjd}):
If to assume that
$\lim_{ {E} \rightarrow { E}_{min}+}
\;\omega ({ E})\stackrel{\rm def}{=} \omega_{0}>0$,
and that
derivatives  $\omega^{(k)}({ E})$, ($k= 0,1,2, \ldots, n$),
are
continuous in
$[ { E}_{min}, \infty)$, (that is
if
for ${ E} > { E}_{min}$ all
$\omega^{(k)}({ E})$
are
continuous
and all the limits
$\lim_{{E} \rightarrow { E}_{min}+}\,\omega^{(k)}({E})$ exist)  and
all
these $\omega^{(k)}({ E})$
are
absolutely integrable functions then \cite{epjd},
\begin{equation}
a(t) \; \begin{array}{c}
          {} \\
          \sim \\
          \scriptstyle{t \rightarrow \infty}
        \end{array}
        \;- \frac{i}{t}\;e^{\textstyle{-\,{i}\,{ E}_{min} t}}\;
        \sum_{k = 0}^{n-1}(-1)^{k} \,\big(\frac{i}{t}\big)^{k}\,\omega^{(k)}_{0}
        = a_{non}(t),
        \label{a-omega}
\end{equation}
 where $\omega^{(k)}_{0}  \stackrel{\rm def}{=} \lim_{{E}\rightarrow {E}_{min}+}
\;\omega^{(k)} ({E})$.

For
a more complicated form of the density $\omega ({E})$
when
$\omega ({ E})$
has the form
\begin{equation}
\omega ({ E}) = ( { E} - { E}_{min})^{\lambda}\;\eta ({ E})\; \in \; L_{1}(-\infty, \infty),
\label{omega-eta}
\end{equation}
where $0 < \lambda < 1$ and it is assumed that $\eta (E_{min}) > 0$ and $\eta^{(k)}({ E})$,
($k= 0,1,\ldots, n$),
exist and they are continuous
in $[{E}_{min}, \infty)$, and  limits \linebreak
$\lim_{{ E} \rightarrow {E}_{min}+}\;\eta^{(k)}({ E})$ exist,
$\lim_{{ E} \rightarrow \infty}\;( { E} - { E}_{min})^{\lambda}\,\eta^{(k)}({ E}) = 0$
for all above mentioned $k$,
one finds that
\begin{eqnarray}
a(t) & \begin{array}{c}
          {} \\
          \sim \\
          \scriptstyle{t \rightarrow \infty}
        \end{array} &
        (-1)\,e^{\textstyle{-{i}{ E}_{min} t}}\;
        \Big[
        \Big(- \frac{i}{t}\Big)^{\lambda + 1}\; \Gamma(\lambda + 1)\;\eta_{0}\; \label{a-eta} \\
        && +\;\lambda\,\Big(- \frac{i}{t}\Big)^{\lambda + 2}\;\Gamma(\lambda + 2)\;\eta_{0}^{(1)}\;+\;\ldots
        \Big]
         = a_{non}(t)
            \nonumber
\end{eqnarray}

\begin{figure}[h!]
\begin{center}
{\includegraphics[height=50mm,width=100mm]{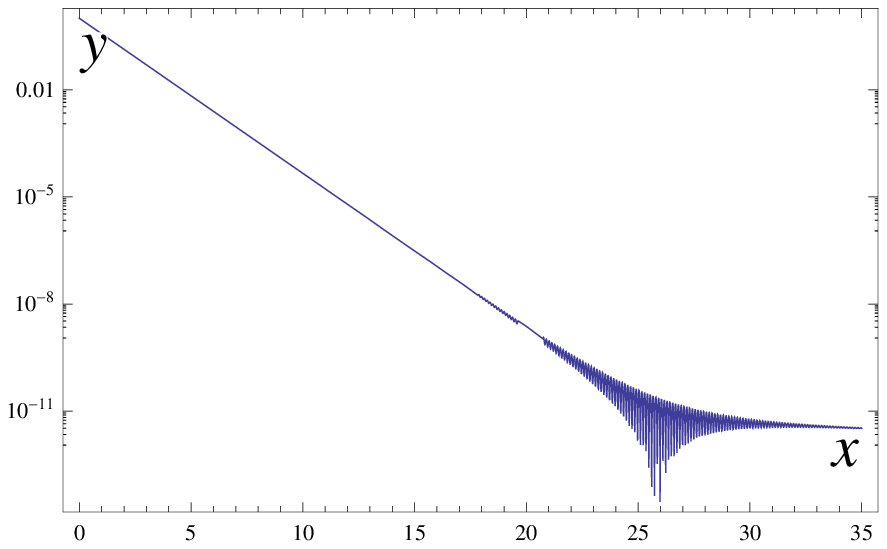}}
\caption{Axes:  $y={\cal P}(t)$  --- the logarithmic scale, $x = t/\tau_{M}$. ${\cal P}(t)$ is the survival probability. The time $t$ is measured as a multiple of the lifetime $\tau_{M}$. The case $({E_{M}^{0}} - E_{min})/ {\it\Gamma_{M}^{0}} = 50$.}
  \label{p2}
\hfill\\
\hfill\\
{\includegraphics[height=50mm,width=100mm]{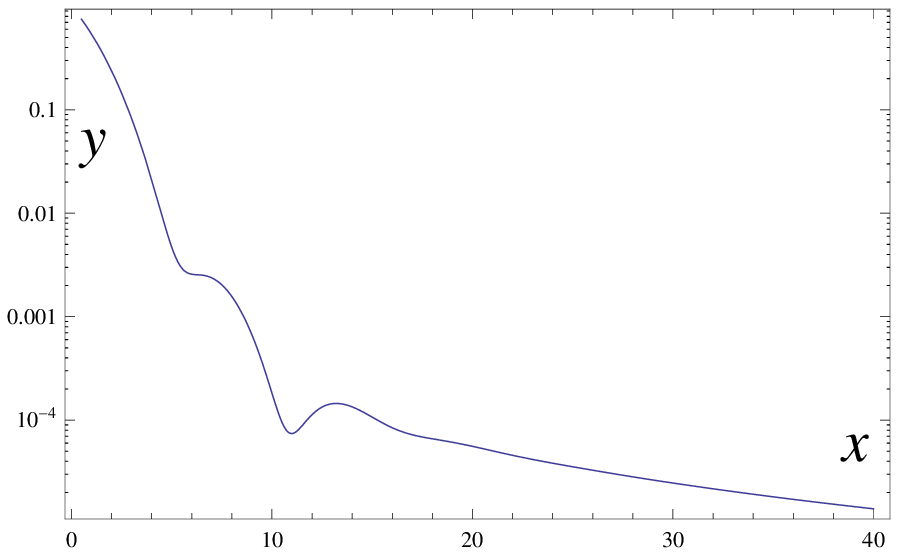}}
\caption{Axes: $y={\cal P}(t)$ --- the logarithmic scale, $x = t/\tau_{M}$. ${\cal P}(t)$ is the survival probability. The time $t$ is measured as a multiple of the lifetime $\tau_{M}$. The case $({E_{M}^{0}} - E_{min})/ {\it\Gamma_{M}^{0}} = 1$.}
  \label{p1}
\end{center}
\vspace{-12pt}
\end{figure}

From (\ref{a-omega}), (\ref{a-eta}) it is seen that asymptotically late time behavior of the survival amplitude $a(t)$ depends rather weakly on a specific form
of the energy density $\omega(E)$. The same concerns a decay curves ${\cal P}(t) = |a(t)|^{2}$.
A typical form of a decay curve, that is the dependence on time $t$ of ${\cal P}(t)$ when $t$ varies from $t = t_{0} =0$ up to  $t > 20 \tau_{M}$ is presented in Fig. (\ref{p2}). The decay curve, which one can observe in the case of  the so--called broad resonances (when $({E_{M}^{0}} - E_{min})/ {\it\Gamma_{M}^{0}} \sim 1$), is presented in Fig (\ref{p1}).
Results presented in these Figures  were obtained for
the  Breit--Wigner   energy distribution function,
\begin{equation}
\omega ({E})
 \equiv \frac{N}{2\pi}\,  {\it\Theta} ( E - E_{min}) \
\frac{{\it\Gamma}_{M}^{0}}{({ E}-{ E}^{0}_{M})^{2} +
({\it\Gamma}_{M}^{0} / {2})^{2}},    \label{omega-BW}
\end{equation}
where $\it\Theta (E) $ is the unit step function.

The crossover time $T$ for this model:
\begin{eqnarray}
{{\it\Gamma}_{M}^{0}\,T}
&\simeq & 8,28 \,+\, 4\,
\ln\,(\frac{{ E}_{M}^{0}\,-
\,{E}_{min}}{{\it\Gamma}_{M}^{0}}) \nonumber \\
&&+\, 2\,\ln\,[8,28 \,+\,4\,\ln\,(\frac{{E}_{M}^{0}\,-
\,{E}_{min}}{{\it\Gamma}_{M}^{0}})\,]\,+\, \ldots \;\;
\label{t-as-3}
\end{eqnarray}
where $({{E}_{M}^{0} - E_{min}}/{{\it\Gamma}_{M}^{0}})\,>\,10$.

\section{Instantaneous energy and instantaneous decay rate}

The amplitude $a(t)$ contains information about
the decay law ${\cal P}(t)$ of the state $|M\rangle$, that
is about the decay rate ${\it\Gamma}_{M}^{0}$ of this state, as well
as the energy ${E}_{M}^{0}$ of the system in this state.
This information can be extracted from $a(t)$. Indeed if
$|M\rangle$ is an unstable (a quasi--stationary) state then
\begin{equation}
a(t)  \cong e^{\textstyle{ - {i}({ E}_{M}^{0} -
\frac{i}{2} {\it\Gamma}_{M}^{0})\,t }}, \;\;(t \sim \tau_{M}). \label{a-q-stat}
\end{equation}
So, there is
\begin{equation}
{E}_{M}^{0} - \frac{i}{2} {\it\Gamma}_{M}^{0} \equiv i
\,\frac{\partial a(t)}{\partial t} \; \frac{1}{a(t)},
\label{E-iG}
\end{equation}
in the case of quasi--stationary states.

The standard interpretation and understanding of the quantum theory
and the related construction of our measuring devices are such that
detecting the energy ${E}_{M}^{0}$ and decay rate
${\it\Gamma}_{M}^{0}$ one is sure that the amplitude $a(t)$ has the
form (\ref{a-q-stat}) and thus that the relation (\ref{E-iG})
occurs. Taking the above into account one can define the "effective
Hamiltonian", $h_{M}$, for the one--dimensional subspace of
states ${\cal H}_{||}$ spanned by the normalized vector
$|M\rangle$ as follows
\begin{equation}
h_{M} \stackrel{\rm def}{=}  i \, \frac{\partial
a(t)}{\partial t} \; \frac{1}{a(t)}. \label{h}
\end{equation}

In general, $h_{M}$ can depend on time $t$, $h_{M}\equiv
h_{M}(t)$. One meets this effective Hamiltonian when one starts
with the Schr\"{o}dinger Equation
for the total state
space ${\cal H}$ and looks for the rigorous evolution equation for
the distinguished subspace of states ${\cal H}_{||} \subset {\cal
H}$.
The equivalent expression for $h_{M} \equiv h_{M}(t)$ has the following form [10]
\begin{equation}
h_{M}(t) \equiv  \frac{\langle M|\mathfrak{H}|M;t\rangle}{\langle M|M;t\rangle}
\stackrel{\rm def}{=} {\cal E}_{M}(t)\,-\,\frac{i}{2}\,\gamma_{M}(t). \label{h2-eq}
\end{equation}
Details can be found in \cite{epjd} and in \cite{cejp}.
Thus,
one finds the following expressions for the
energy and the decay rate of the system in the state $|M\rangle$
under considerations, to be more precise for
the instantaneous energy ${\cal E}_{M}(t)$ and the instantaneous decay rate,
$\gamma_{M}(t)$,
\begin{eqnarray}
{\cal E}_{M}&\equiv& {\cal E}_{M}(t) = \Re\,(h_{M}(t)),
\label{E(t)}\\
\gamma_{M} &\equiv& \gamma_{M}(t) = -\,2\,\Im\,(h_{M}(t)),
\label{G(t)}
\end{eqnarray}
where $\Re\,(z)$ and $\Im\,(z)$ denote the real and imaginary parts
of $z$ respectively.

Using (\ref{h}) and (\ref{E(t)}), (\ref{G(t)}) one can find that
\begin{eqnarray}
{\cal E}_{M} (0) &=& \langle M |\mathfrak{H}| M \rangle, \\
{\cal E}_{M} (t \sim \tau_{M}) & \simeq & { E}_{M}^{0} \;\;\neq \;\; {\cal E}_{M} (0),\\
\gamma_{M}(0) &=& 0,\\
\gamma_{M}(t \sim \tau_{M}) &\simeq & {\it\Gamma}_{M}^{0}.
\end{eqnarray}
So, there is ${\cal E}_{M}(t)= E_{M}^{0}$ at the canonical decay time.

Starting from the  asymptotic expressions (\ref{a-omega}) and (\ref{a-eta}) for $a(t)$ and using (\ref{h})
after some algebra one finds for  times $t \gg T$  that
\begin{equation}
{h_{M}(t)\vline}_{\,t \rightarrow \infty} \simeq { E}_{min} + (-\,\frac{i}{t})\,c_{1} \,
+\,(-\,\frac{i }{t})^{2}\,c_{2} \,+\,\ldots, \label{h-infty-gen}
\end{equation}
where $ c_{i} = c_{i}^{\ast},\;\;i = 1,2,\ldots$; (coefficients $c_{i}$ depend on  $\omega (E)$).
This  last relation means that
\begin{eqnarray}
{\cal E}_{M}(t) &\simeq & E_{min} \, -  \,\frac{c_{2}}{t^{2}} \ldots, \;\;\;({\rm for}
\;\;t \gg T), \label{E(t)}\\
\gamma_{M}(t) &\simeq & 2\,\frac{c_{1}}{t} \,+\ldots, \;\;\;({\rm for}
\;\;t \gg T), \label{G(t)}
\end{eqnarray}
These properties take place for  all unstable states which survived up to times $t \gg T$.

Note that from (\ref{E(t)}) it follows that $\lim_{t \rightarrow \infty}\, {\cal E}_{M}(t) = E_{min}$.

For the most general form (\ref{omega-eta}) of the density $\omega (E)$ (i. e. for $a(t)$ having the asymptotic
form given by (\ref{a-eta}) )  we have
\begin{equation}
c_{1} = \lambda + 1, \;\;\;\;c_{2} =  (\lambda + 1)\,\frac{\eta^{(1)}(E_{min})}{\eta (E_{min})} . \label{c-i}
\end{equation}
The energy densities $\omega (E)$ considered in quantum mechanics and in quantum field theory can be described by
$\omega (E)$ of the form  (\ref{omega-eta}), eg.  quantum field theory models correspond with $\lambda = \frac{1}{2}$.

The average energy measured at some time interval $(t_{1},t_{2})$ (with $t_{1}, t_{2} \gg T$)
equals
\begin{equation}
\overline{{\cal E}_{M}(t)} = \frac{1}{t_{2} - t_{1}}\,\int_{t_{1}}^{t_{2}}\,{\cal E}_{M}(t)\,dt \simeq
E_{min}\, -\, \frac{c_{2}}{t_{1}\,t_{2}} + \ldots , \label{E-sr}
\end{equation}

\begin{figure}[h!]
\begin{center}
{\includegraphics[height=50mm,width=100mm]{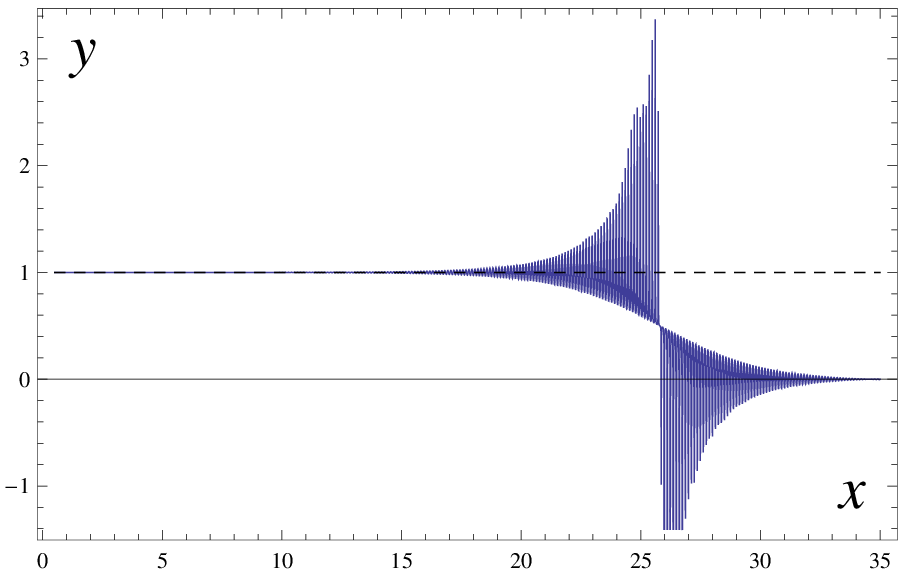}}
\caption{Axes: $y= ({\cal E}_{M}(t) - E_{min}) /  (E_{M}^{0} - E_{min})$, $x= t/\tau_{M}$. The difference  of energies $({\cal E}_{M}(t) - E_{min})$ is measured as a multiple of the difference $(E_{M}^{0} - E_{min})$. The case  $({E_{M}^{0}} - E_{min})/ {\it\Gamma_{M}^{0}} = 50$}.
  \label{E1}
\hfill\\
\hfill\\
{\includegraphics[height=50mm,width=100mm]{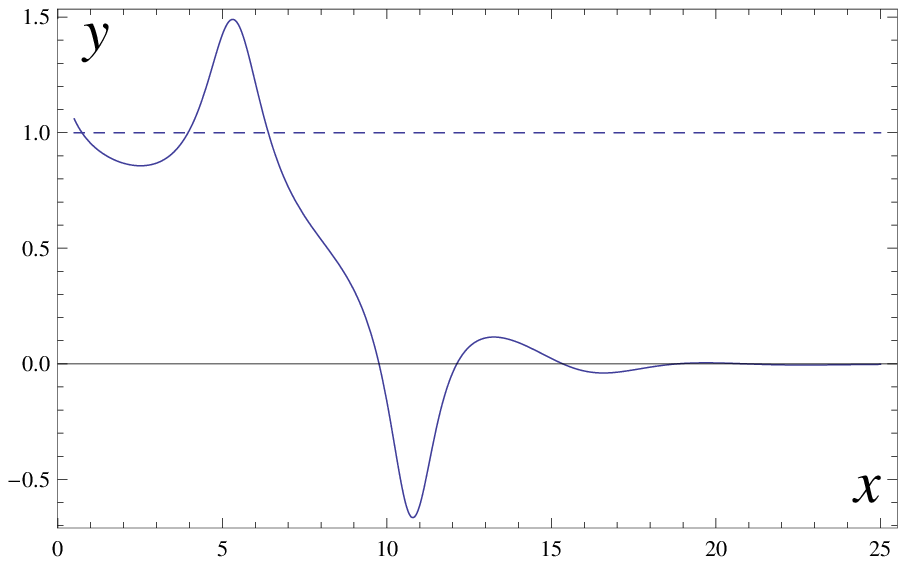}}
\caption{Axes:  $y= ({\cal E}_{M}(t) - E_{min}) /  (E_{M}^{0} - E_{min})$, $x= t/\tau_{M}$. The difference  of energies $({\cal E}_{M}(t) - E_{min})$ is measured as a multiple of the difference $(E_{M}^{0} - E_{min})$. The case  $({E_{M}^{0}} - E_{min})/ {\it\Gamma_{M}^{0}} = 1$. }
  \label{E2}
\end{center}
\vspace{-12pt}
\end{figure}

A general form of  $({\cal E}_{M}(t) - E_{min}) /  (E_{M}^{0} - E_{min})$  as a function of time $t$ varying from $t = t_{0} =0$ up to $ t > T$ is presented
in  Figs (\ref{E1}), (\ref{E2}). These results were obtained for the model considered in  the previous Section and correspond with Figs (\ref{p2}), (\ref{p1}).

\section{Cosmological applications}

Krauss and Dent in their paper \cite{Krauss} mentioned earlier
made a hypothesis
that some false vacuum regions do survive well up to the time $T$ or  later.
Let $|M\rangle = | 0\rangle^{false}$,
be a false, $|0\rangle^{true}$ -- a true, vacuum states and   $E^{false}_{0}$ be the energy of a state corresponding to the false vacuum measured at the canonical decay time
and  $E^{true}_{0}$ be the energy of true vacuum (i.e. the true ground state of the system).
As it is seen from the results presented in previous Section, the problem is that the energy of those false
vacuum regions which survived up to $T$ and much later differs from $E^{false}_{0}$  \cite{PRL-2011}.

Now, if  one assumes that $E^{true}_{0} \equiv E_{min}$ and $E_{0}^{false} =  E_{M}^{0}$ and takes into account results of the previous Section (including those in Figs (\ref{E1}), (\ref{E2}))
then one can conclude that the energy of the system in  of  false vacuum state has the following general properties:

\begin{equation}
{\cal E}^{false}_{0}(t) = E^{true}_{0}  + \Delta E \;\cdot\;\Phi(t),
\label{E-false-(t)}
\end{equation}
where $\Delta E = E_{0}^{false} - E^{true}_{0}$ and $\Phi (t) = \frac{{\cal E}^{false}_{0}(t) - E^{true}_{0}}{\Delta E}\simeq 1$ for $t \sim \tau_{0}^{false} < T$. $\Phi (t)$ is a fluctuating function of $t$   at $t \sim T$ (see Figs (3), (4)) and   $ \Phi (t) \propto \frac{1}{t^{2}}$ for $t \gg T$.

At asymptotically late times,
$t \gg T$, one finds that
\begin{equation}
{\cal E}^{false}_{0}(t) \simeq E^{true}_{0}  - \frac{c_{2}}{t^{2}}\ldots \;\; \neq\,E^{false}_{0},
\label{E-false-infty}
\end{equation}
where $c_{2} = c_{2}^{\ast}$ and it can be positive or negative depending on the model considered.
Similarly
\begin{equation}
\gamma^{false}_{0}(t) \simeq   + 2\,\frac{c_{1}}{t}\ldots \;\; \;({\rm for} \;\;t \gg T).
\label{G-false-infty}
\end{equation}
Two last properties of the false vacuum states mean that
\begin{equation}
{\cal E}^{false}_{0}(t) \rightarrow E^{true}_{0} \;\;{\rm and}\;\;
\gamma^{false}_{0}(t) \rightarrow 0 \;\;{\rm as}\;\; t\rightarrow \infty. \label{E-false-lim}
\end{equation}

Going from quantum mechanics to quantum field theory one should take into account  among others a volume factors so that survival probabilities per unit volume per unit time should be considered. The standard false vacuum decay calculations shows that the same volume factors should appear in both early and late time decay rate estimations (see Krauss and Dent \cite{Krauss} ). This means that the calculations of cross--over time $T$ can be applied to survival probabilities per unit
volume.  For the same reasons  within the quantum field theory the quantity  ${\cal E}_{M}(t)$ can be replaced by  the energy per unit volume $\rho_{M}(t) = {\cal E}_{M}(t)/V$ because these volume factors $V$  appear in the numerator and denominator of the formula (\ref{h}) for $h_{M}(t)$.   This conclusion seems to hold when considering the energy ${\cal E}_{0}^{false}(t)$ of the system in false vacuum state $|0\rangle^{false}$ because Universe is assumed to be homogeneous and isotropic at suitably large scales. So at such scales to a sufficiently good accuracy we can extract properties of the energy density $\rho_{0}^{false}$ of the system in the false vacuum state $|0\rangle^{false}$ from properties of the energy ${\cal E}_{0}^{false}(t)$ of the system in this state defining $\rho_{0}^{false}(t)$ as $\rho_{0}^{false}(t) = {\cal E}_{0}^{false}(t)/V$. Thus one can conclude from (\ref{E-false-(t)}) and from (\ref{E-false-infty}) that
the energy density $\rho^{false}_{0}(t)$ in the unstable false vacuum state  has the following properties as a function of time $t$,
\begin{equation}
{\rho}^{false}_{0}(t) = \rho^{true}_{0}  + D\,\cdot \,F(t),
\label{rho-false-(t)}
\end{equation}
where $D=D^{\ast} $, $\rho^{true}_{0} \equiv \rho_{0}^{bare}$, $F(t) \simeq 1$ for $t \sim \tau_{0}^{false}$, $F(t) \propto \Phi(t) $ is fluctuating at $t \sim T$ and $F(t) \sim 1/t^{2}$ at $t \gg T$  and the sign of $D$ depends on the model considered.

Similarly the asymptotically late time behavior of  the energy density $\rho^{false}_{0}(t)$ is given by the following relation \cite{aip}
\begin{equation}
{\rho}^{false}_{0}(t) \simeq \rho^{true}_{0}  - \frac{d_{2}}{t^{2}}\ldots ,\;\;\; {\rm for}\;\;\; t \gg T,
\label{rho-false-infty-2}
\end{equation}
(where $d_{2}=d_{2}^{\ast}$, $\rho^{true}_{0} \equiv \rho_{0}^{bare}$.

The standard relation is
\begin{equation}
\rho_{0} \equiv \rho^{true}_{0} =  \frac{\Lambda_{0}}{8 \pi G}, \label{Lambda-rho}
\end{equation}
where $\Lambda_{0} \equiv \Lambda^{bare}$ is the bare cosmological constant. Therefore conclusions (\ref{rho-false-(t)}), (\ref{rho-false-infty-2}) hold for $\Lambda = \Lambda (t)$ too.

\section{ Final Remarks}

The basic physical factor forcing the wave function  $|M;t\rangle$ and thus the  amplitude $a(t)$
to exhibit inverse power law behavior at $t \gg T$ is a boundedness from below of the spectrum  $\sigma (\mathfrak{H})$ of the total Hamiltonian $\mathfrak{H}$ of the system under considerations. This means
that if this condition takes place and
\begin{equation}
\int _{-\infty}^{+\infty}\omega(E)\,dE\,< \,\infty, \label{omega-absolute}
\end{equation}
then all  properties
of $a(t)$, including a form of the time--dependence at  $t \gg T$, are the  mathematical consequence of them both.
The same applies by (\ref{h}) to properties of $h_{M}(t)$ and concerns
the asymptotic form of $h_{M}(t)$ and
thus of ${\cal E}_{M}(t)$  and $\gamma_{M}(t)$ at $t \gg T$.
(Note that properties of $a(t)$ and $h_{M}(t)$ discussed above
do not take place when  $\sigma(\mathfrak{H}) = (-\infty, + \infty)$).
So the late time behavior of the energy density ${\rho}^{false}_{0}(t)$  in the false vacuum state (\ref{rho-false-(t)}), (\ref{rho-false-infty-2}),
is the pure quantum effect following from the basic principles of the quantum
theory (see relations (\ref{a-eta}), (\ref{h}), (\ref{E(t)})).

The late time properties of the energy of the unstable false vacuum state discussed in the previous Section give a strong support for cosmological models using:
\[
\rho_{0} (t) = \rho_{0}^{bare} + \frac{A_{0}}{t^{2}}, \;\;\;{\rm or}\;{\rm equivalently,}\;\;\;\;
\Lambda (t) = \Lambda^{bare} + \frac{B_{0}}{t^{2}},
\]
where $A_{0}, B_{0}$ are real and can be positive or negative depending on the model considered,
and for models with
\[
\rho_{0} (t) = \rho_{0}^{bare} + A_{1}\,H^{2} \;\;\;{\rm or} \;{\rm equivalently},\;\;\;\;
\Lambda (t) = \Lambda^{bare} + B_{1}\,H^{2},
\]
where $A_{1}, B_{1}$ are real and $H$ is the current Hubble constant. (There is \linebreak $H   \propto \frac{1}{t}$).

Cosmologies using such parameters are consistent with the quantum theoretical treatment of unstable vacua.

There are many open problems in the approach discussed above. For example:
Properties of the energy ${\cal E}^{false}_{0}(t)$ are determined by the form of the energy density $\omega (E)$
(The sign of $\eta^{(1)}(E_{min})$  and thus the sign of  $c_{2}$ in the formula (\ref{c-i}) depends on the form of $\eta(E)$ and thus of $\omega(E)$):
It is necessary to find at least an approximate  form of  $\omega(E)$ for false vacuum states. An another problem:
As it is seen from Figs (3), (4), the energy ${\cal E}^{false}_{0}(t)$ of the unstable false vacuum state should fluctuate at transition times $t \sim T$. This means that ${\rho}^{false}_{0}(t)$ and $\Lambda(t)$ should also fluctuate at these times.
The question is: What are possible consequences of  this effect?
Yet one other problem: If the vacuum  in  Universe is unstable (or even metastable)  why  our Universe still exists?


\begin{thebibliography}{6}
\bibitem{Coleman}
S. Coleman, Phys. Rev. D 15, 2929 (1977);  C.G. Callan
and S. Coleman, Phys. Rev. D 16, 1762 (1977).
\bibitem{Coleman2}
S. Coleman and F. de Lucia, Phys. Rev. D 21, 3305 (1980).
\bibitem{Krauss}
L. M. Krauss, J. Dent, Phys. Rev. Lett., \textbf{100}, 171301 (2008).
\bibitem{Spencer}
 A. Kobakhidze, A. Spencer--Smith,  Phys. Lett. {\bf B 722}, 130, (2013).
 \bibitem{Degrassi}
 G. Degrassi, {\em et al.},JHEP 1208 (2012) 098.
 \bibitem{Elias}
J. Elias--Miro, {\em et al.}, Phys. Lett. {\bf B 709}, 222, (2012).
\bibitem{Wei}
Wei Chao, {\em et al.}, Phys. Rev. {\bf D 86}, 113017, (2012).
\bibitem{Khalfin}
L. A. Khalfin, Zh. Eksp. Teor. Fiz. {\bf 33}, 1371 (1957),
[Sov. Phys. JETP {\bf 6}, 1053 (1958)].
\bibitem{epjd}
K. Urbanowski, {\em Eur. Phys. J. D}, {\bf 54}, (2009);
 (DOI: 10.1140/epjd/e2009-00165-x).
 \bibitem{cejp}
 K. Urbanowski, Cent. Eur. J. Phys. {\bf 7}, (2009), (see also references one can find therein).
 \bibitem{PRL-2011}
 K. Urbanowski, {\em Phys. Rev. Lett.}, {\bf 107}, 209001 (2011),
(see also references one can find therein).
\cite{aip}
\bibitem{aip}
K. Urbanowski, M. Szyd³owski,
{\em AIP Conf. Proc.} {\bf 1514}, 143 (2013); doi: 10.1063/1.4791743 .
\end{thebibliography}
\end{document}